\documentclass[twocolumn,english,superscriptaddress,notitlepage]{revtex4-2}
\usepackage[T1]{fontenc}
\usepackage[latin9]{inputenc}
\usepackage{graphicx}
\usepackage{amsmath}  

\makeatletter
\usepackage{babel}
\usepackage{color}
\usepackage[colorlinks = true,
linkcolor = blue,
urlcolor  = blue,
citecolor = blue,
anchorcolor = blue]{hyperref}

\usepackage{tikz}
\usetikzlibrary{patterns,snakes}

\makeatother

\mathchardef\mhyphen="2D 

\begin{document}

\title{Explaining the extra crystal field mode in $A$Ce$X_2$}

\author{Allen O. Scheie}
\email{scheie@lanl.gov}
\affiliation{MPA-Q, Los Alamos National Laboratory, Los Alamos, New Mexico 87545}

\author{Sabrina J. Li}
\affiliation{Computational Physics Division, Los Alamos National Laboratory, Los Alamos, New Mexico 87545}
\affiliation{Center for Nonlinear Studies, Los Alamos National Laboratory, Los Alamos, New Mexico 87545}

\author{Stephen D. Wilson}
\affiliation{Materials Department and California Nanosystems Institute, University of California Santa Barbara, Santa Barbara, California 93106}

\author{Daniel A. Rehn}
\affiliation{Computational Physics Division, Los Alamos National Laboratory, Los Alamos, New Mexico 87545}

\date{\today}

\begin{abstract}
    A growing list of Ce-based magnets have shown an extra and heretofore unexplained crystal electric field (CEF) mode at high energies.  
    We describe a process whereby an optical phonon can produce a split CEF mode well above the phonon energy. We use density functional theory and point-charge model calculations to estimate the phonon distortions and coupling to model this effect in KCeO$_2$, showing that it accounts for the extra CEF mode observed. This mechanism is generic, and may explain the extra modes observed on a variety of Ce$^{3+}$ compounds. 
\end{abstract}

\maketitle

\section{Introduction}

Rare earth ion crystal electric field (CEF) excited levels are known to couple to phonons to form vibronic bound states that are hybrid CEF phonon excitations  \cite{Adroja_2012,vcermak2019magnetoelastic,DELAFUENTE2021167541,Anand_2021,Princep_2013_CEF,Scheie2024_KYS,pocs2025heat}. 
This effect is well-understood when the phonons and CEF modes are close in energy. However, a series of Ce-based magnetic compounds have recently been reported which have anomalous extra CEF modes well above the phonon energies. This is clearly seen in KCeO$_2$ \cite{Bordelon_2021}, KCeS$_2$ \cite{Bastien_2020}, and RbCe$X_2$ ($X=$O, S, Se, Te) \cite{Ortiz_2022}; and to a lesser degree in other Ce$^{3+}$ compounds \cite{Gaudet_2019_CZO,gao2019experimental,Shin_2020}.  

This is a conundrum for Ce$^{3+}$, which has a $J=5/2$ ground state and is constrained by Kramer's theorem to have three doublets in zero field---and thus only two excited CEF levels. Consequently, when a \textit{third} level is observed which is at a similar energy to the other CEF levels, is well above the optical phonon energies, follows the Ce$^{3+}$ form factor, and shows no dispersion---in other words, looks spectroscopically exactly like an excited CEF mode---it is \textit{prima facie} unclear where this could come from (and furthermore makes the CEF Hamiltonian impossible to fit accurately).  
Adding to the puzzle is that this behavior is not universal: a slight structural modification from $D_{3d}$ symmetry in KCeO$_2$ to $D_{2d}$ symmetry in NaCeO$_2$ apparently eliminates the extra mode \cite{Bordelon_2021_NCO,Bhattacharyya_2022}.

Here we propose a mechanism to explain this puzzling behavior as a vibronic mode formed from a higher harmonic of a CEF mode coupled to an optical phonon. This can split a single CEF mode into what spectroscopically appears as two CEF excitations. 
Focusing on KCeO$_2$, we identify the specific optical phonon modes which produce this behavior, and then generalize the mechanism to other Ce$^{3+}$ materials. 

\section{Calculation methods and results}

\begin{figure}
    \centering
    \includegraphics[width=0.85\linewidth]{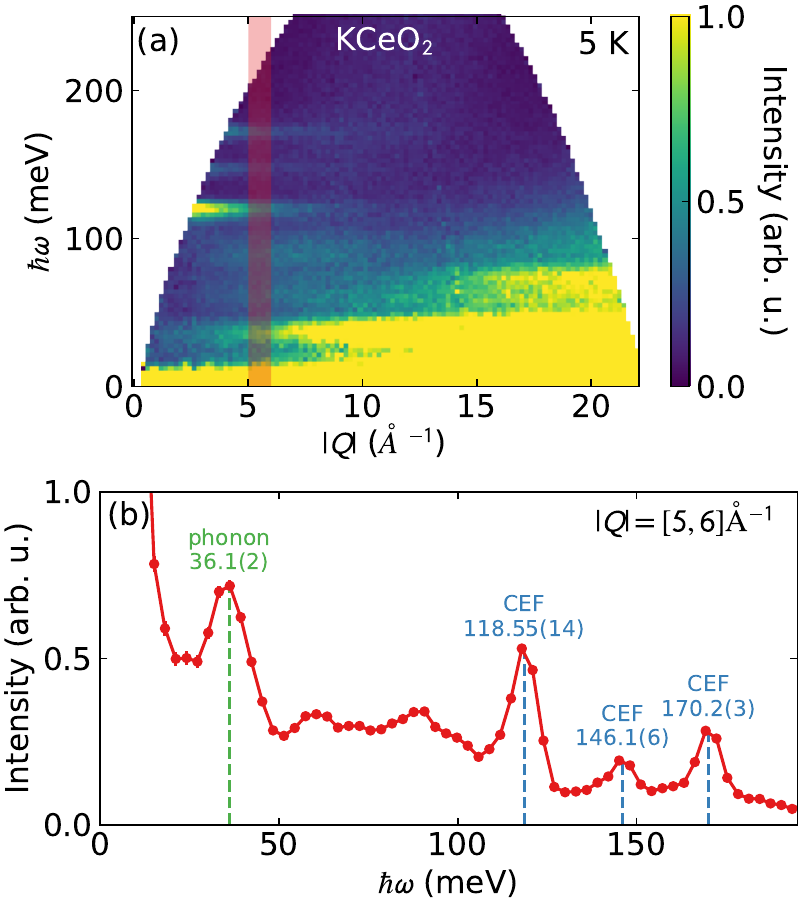}
    \caption{KCeO$_2$ inelastic neutron scattering measured with ARCS with $E_i=300$~meV neutrons. Panel (a) shows the full spectrum, panel (b) shows a constant $Q$ cut between 5 and 6~\AA$^{-1}$ [indicated by the vertical red bar in panel (a)]. There is a clear optical phonon at 36.1 meV which shows increasing intensity with $|Q|$, and then three CEF-like modes which show decreasing intensity with $|Q|$.}
    \label{fig:NeutronData}
\end{figure}

KCeO$_2$ is a candidate material for quantum spin liquid behavior, and has three observed CEF modes measured with the ARCS spectrometer \cite{ARCS} at the ORNL Spallation Neutron Source \cite{mason2006spallation} and reported in Ref. \cite{Bordelon_2021}. The data are shown in Fig. \ref{fig:NeutronData}. Three CEF modes---distinguished by their following a magnetic form factor, decreasing intensity with $|Q|$---are found at $\hbar \omega = 118.55(14)$~meV, 146.1(6)~meV, and 170.2(3)~meV. 
The phonon scattering is at lower energies, seen as intense modes that increase intensity with $|Q|$. 

To obtain a clearer picture of the vibrational spectra of KCeO$_2$, we used density functional theory (DFT) to calculate phonons. 
Phonons were calculated using the VASP code, version 5.4.4~\cite{kresse1993ab, kresse1996efficient, kresse1996efficiency}.  The PAW method~\cite{blochl1994projector,kresse1999ultrasoft} was used for all calculations, treating Ce with valence configuration $5s^25p^66s^25d^{1}$, K with $3p^64s^1$, and O with $2s^22p^4$. The PBE~\cite{perdew1996generalized} exchange-correlation functional was used and all calculations were performed using a Gamma-centered Monkhorst-Pack~\cite{monkhorst1976special} mesh and Gaussian smearing. Phonons were calculated using the finite displacement as implemented in the Phonopy software~\cite{togo2023first}. We calculated phonons for different supercells of the primitive cell (4 atom) to test convergence with respect to cell size (see Supplemental Material). We found that a $4\times4\times4$ supercell of the primitive cell (256 atoms) were needed to obtain fully converged phonon dispersions. An energy cutoff of 600 eV and a k-mesh of size $3\times3\times3$ were found to be sufficient to obtain converged forces and phonons at this supercell size. Note also that we include the effects of LO-TO splitting through calculation of Born effective charges. LO-TO splitting is significant for high energy optical modes at the $\Gamma$ point. The primitive unit cell and path through the Brillouin zone are obtained using the convention of the AFLOW software~\cite{setyawan2010high,curtarolo2012aflow} (see Supplemental Material for more information). Note that phonons for NaCeO$_2$ are also provided in the Supplemental Material.

\begin{figure}
    \centering
    \includegraphics[width=\linewidth]{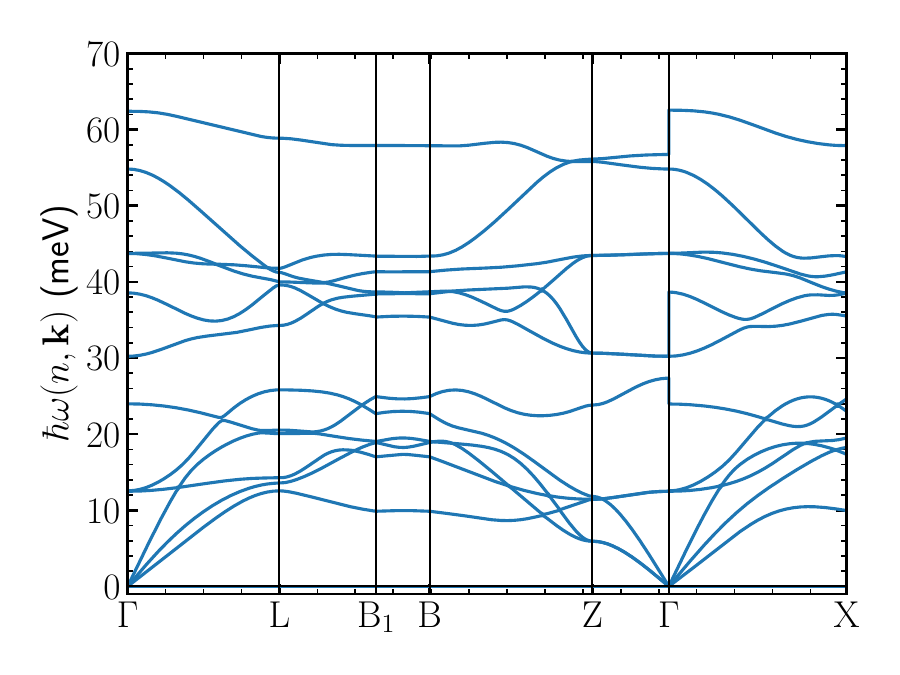}
    \caption{KCeO$_2$ phonon dispersions calculated with density functional theory (DFT), plotted along high symmetry directions. LO-TO splitting is included in these calculations.}
    \label{fig:PhononBandStructure}
\end{figure}

The calculated phonon dispersions are plotted in Fig.~\ref{fig:PhononBandStructure}.
One can see immediately that all the phonon bands are far below  118~meV, the lowest CEF mode. Thus we can rule out a direct vibronic coupling between phonons and CEF modes.

However, this does not mean the CEF coupling to phonons has no effect. To demonstrate this, we  calculate the effect of CEF phonon coupling via the point charge model. This method models the ligands surrounding a magnetic ion as Coulombic point charges \cite{Hutchings1964} which, although it neglects extended orbitals and charge screening, is often a reasonably accurate approximation for static CEF Hamiltonians \cite{EDVARDSSON1998230}. 

We construct a point charge model for KCeO$_2$ using \textit{PyCrystalField} software \cite{PyCrystalField}, and adjust the ligand effective charges to account for charge screening and extended orbitals so that the CEF levels roughly match experiment (calculated $\hbar \omega =126.6$~meV and 171.1~meV, from an effective O charge $-1.95e$ and an effective neighboring Ce charge $1.0e$, giving CEF parameters $B_2^0 = 4.376$~meV, $B_4^0 = -0.201$~meV, $B_4^3 = 7.642$~meV). 
We then distort the ligands according to the calculated phonon modes (see Appendix \ref{app:details}), and recalculate the CEF Hamiltonian from the point charge model. We write the  phonon distortion $\mu$ in Stevens operators $\mathcal{O}_n^m$ \cite{Stevens1952} as the difference between the un-distorted CEF Hamiltonian and the distorted CEF Hamiltonian $\mathcal{O}_{\mu} = \sum_{n,m} \left[ (B_n^m)_0 - (B_n^m)_{distorted} \right] \mathcal{O}_n^m$. 
We then write the full CEF-phonon Hamiltonian as 
\begin{equation}
    \mathcal{H} = \mathcal{H}_{CEF} + \sum_\mu \left[ \hbar \omega_\mu \left(a_{\mu}^{\dagger} a_{\mu} + \frac{1}{2} \right) + (a_{\mu} + a_{\mu}^{\dagger} ) \mathcal{O}_{\mu} \right] 
    \label{eq:CEF-phonon}
\end{equation}
\cite{Thalameier_1982,Thalmeier_1984} 
where $\mathcal{H}_{CEF}$ is the un-distorted CEF Hamiltonian, $a_{\mu}$ and $a_{\mu}^{\dagger}$ are the phonon creation and annihilation operators of phonon $\mu$ at energy $\hbar \omega_\mu$, and $\mathcal{O}_{\mu}$ is the phonon distortion operator defined above. 

\begin{figure}
    \centering
    \includegraphics[width=1.03\linewidth]{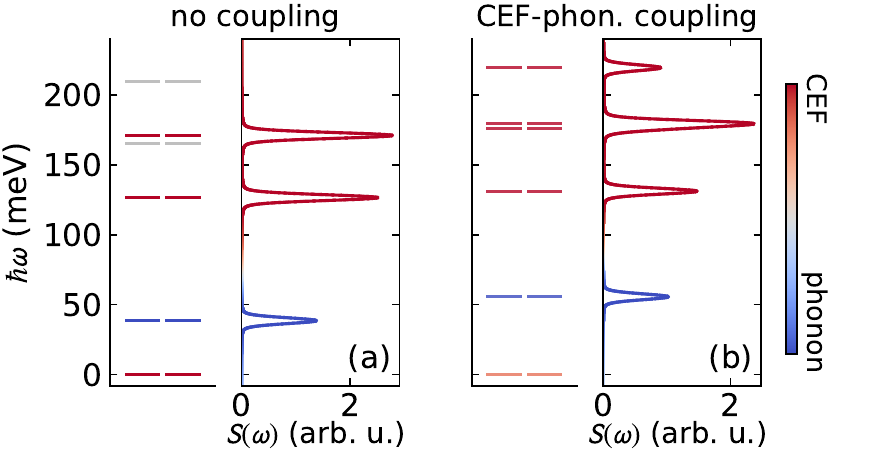}
    \caption{Example plot of calculated phonon mode 8 at ${\bf Q} = (0.02,0,0)$ coupled to the KCeO$_2$ CEF Hamiltonian, with the distortion amplitude multiplied by 3 to exaggerate the effect. (a) Eigenvalues (left) and intensities (right) of the Hamiltonian without CEF-phonon coupling. (b) Eigenvalues and intensities of the Hamiltonian with CEF-phonon coupling. The color indicates the CEF ($\langle 0 | J_{\pm} | n \rangle$) or phonon ($\langle 0 | a^\dagger | n \rangle$) character of the eigenstate. Note that without the CEF coupling, some eigenstates have zero intensity (shown in grey) as they have no overlap with the ground state.}
    \label{fig:Example}
\end{figure}

KCeO$_2$ has 12 phonon modes, and it is computationally prohibitive to diagonalize all phonons simultaneously with the CEF Hamiltonian. Therefore we compute the eigenspectrum of Eq. \ref{eq:CEF-phonon} for each phonon mode individually across several high-symmetry directions in reciprocal space. We then calculate the $|\langle 0 | J_{\pm} | n \rangle|^2$ overlap to each excited state from the ground state doublet to determine the CEF spectrum under the phonon coupling (which would correspond to the magnetic neutron spectrum). 
 The full results are shown in Appendix \ref{app:details}, but an example phonon distortion is shown in Fig. \ref{fig:Example} which shows the eigenspectrum with and without the CEF-phonon coupling.  

\section{Discussion}

Figure \ref{fig:Example} shows that a phonon mode well below the CEF levels can produce a substantial effect on CEF states. This is because there are higher harmonics in the eigenspectrum which are combinations of the phonon energy and CEF energy. In the absence of CEF-phonon coupling, these have zero intensity. However when the coupling between the CEF and phonons becomes nonzero, they can begin to pick up intensity, especially when a higher harmonic is close in energy to another CEF level. (Note however that this requires sufficiently strong coupling---the phonon distortions in Fig. \ref{fig:Example} were multiplied by 3 from the DFT calculated values to exaggerate the effect.)

Appendix \ref{app:details} shows that only some phonon modes affect the CEF eigenspectrum, but those that do have an effect can produce a noticeable shift or splitting of the CEF peaks. 
Modes 4, 5, 7, and 8 in particular show substantial splitting of the CEF levels near $\Gamma$ ($Q=0$). These modes correspond to a phonon distortion where the oxygen ligand cage shifts in the $ab$ plane. Table \ref{tab:KCO_phononOperators} in Appendix \ref{app:details} shows that  $B_2^1$ dominates the CEF-phonon operators of these phonon distortions. 

\begin{figure}
    \centering
    \includegraphics[width=0.95\linewidth]{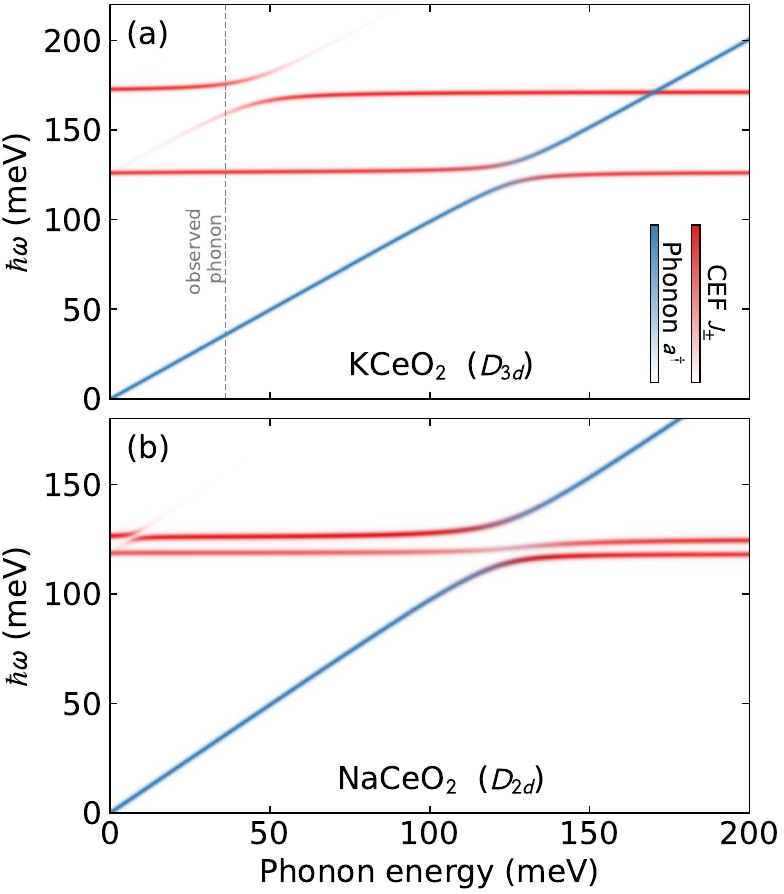}
    \caption{CEF and phonon intensities of a $B_2^1$ phonon distortion as a function of the phonon energy, for KCeO$_2$ (a) and NaCeO$_2$ (b). In both cases the phonon couples to the highest CEF eigenstate through a CEF + phonon harmonic, and produces substantial splitting. 
    This effect is present in both models, but is more pronounced for KCeO$_2$. Furthermore, the experimental optical phonon energy of 36 meV in KCeO$_2$ (shown by the grey dashed line) is within range of where the higher harmonic interferes with the highest eigenstate, producing three observed CEF modes.}
    \label{fig:ToyModel}
\end{figure}

To illustrate this effect more clearly, we construct a simplified toy model of KCeO$_2$ including a phonon distortion of solely $B_2^1 = 3.5$~meV. 
The CEF and phonon spectrum is plotted in Fig. \ref{fig:ToyModel} as the phonon frequency is varied from zero to above the highest CEF level. 
Interestingly, when the combined energy of the lowest CEF excited state and the optical phonon approaches the energy higher CEF level, this produces a noticeable  hybridization with this higher harmonic mode, splitting the spectroscopic peaks. Then as the phonon energy increases further the lower CEF mode hybridizes with the phonon as it crosses the mode energy directly, but does not interfere with the higher CEF mode directly.

We repeated this toy model calculation for NaCeO$_2$ \cite{Bordelon_2021_NCO}, which has a different space group and different Ce$^{3+}$ point group symmetry. Although the space group and Hamiltonian are different, the effect is still visible (though less pronounced, and at a phonon energy lower than the experimental optical phonons). So the possibility of a CEF-phonon harmonic splitting a CEF mode is a generic feature which is not restricted to a particular point group symmetry, but rather depends on the details of the CEF Hamiltonian and the phonon energies.

And thus we identify a mechanism for the three CEF modes observed in Ce materials. In KCeO$_2$, a strong optical phonon signal is measured at $36.1(2)$~meV (Fig. \ref{fig:NeutronData}) \cite{Bordelon_2021}, which is close to the DFT-calculated modes 7-8. This mode has approximately the energy and distortion to create a splitting in the higher CEF mode from a harmonic of the lower 119~meV CEF mode and the 35~meV phonon. 

\begin{figure}
    \centering
    \includegraphics[width=0.9\linewidth]{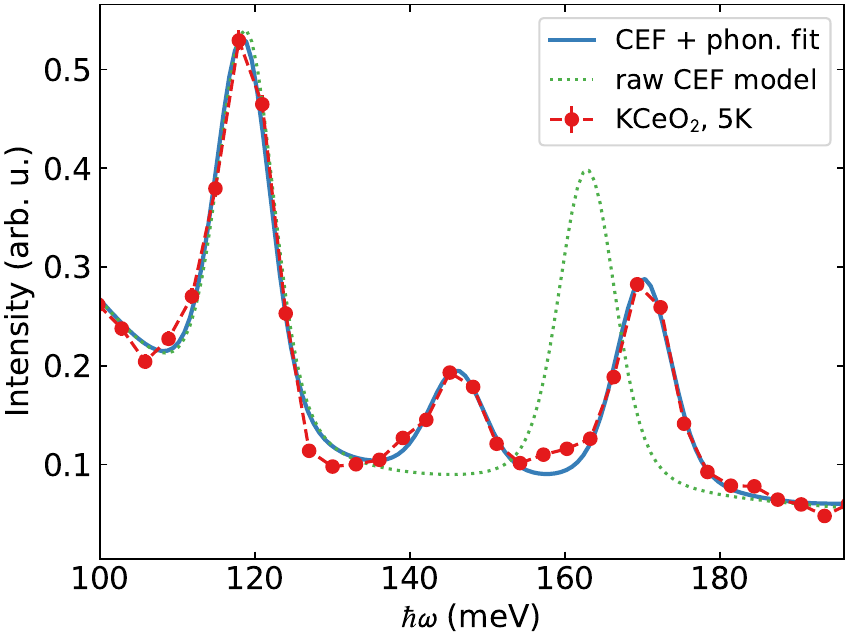}
    \caption{CEF + phonon model fit to KCeO$_2$ inelastic neutron scattering data \cite{Bordelon_2021} assuming a $B_2^1$ CEF-phonon coupling. The fitted parameters are listed in the main text, but the model matches the experimental data remarkably well. The green dashed line shows the fitted CEF model with the phonon coupling removed---what would be observed in the absence of phonons.}
    \label{fig:refit}
\end{figure}

To demonstrate this effect quantitatively, we re-fit the KCeO$_2$ inelastic neutron data from Ref. \cite{Bordelon_2021}, assuming a $B_2^1$ CEF-phonon coupling, and fitting the three nonzero single-ion CEF parameters allowed for $D_{3d}$ symmetry: $B_2^0$, $B_4^0$, and $B_4^3$. The results are shown in Fig. \ref{fig:refit}. 
The fit including the phonon coupling matches the data extremely well. The fitted phonon energy is $(33.2 \pm 1.4)$~meV, the fitted CEF-coupling is $B_2^1=5.1(4)$~meV, and the fitted static CEF parameters are $B_2^0= 5.6(4)$~meV, $B_4^0 = -0.200(5)$~meV, and $B_4^3 = 6.2(4)$~meV, yielding a CEF ground state $| \psi_{\pm} \rangle = \pm 0.39(3) | \pm \frac{5}{2} \rangle - 0.919(12) | \pm \frac{1}{2} \rangle$. 
Uncertainties are estimated using the method in Ref. \cite{Scheie-CEF-uncertainty}. 
The new fitted ground state is close to but slightly different from $| \psi_{\pm} \rangle = \pm 0.473 | \pm \frac{5}{2} \rangle - 0.881 | \pm \frac{1}{2} \rangle$ reported in Ref. \cite{Bordelon_2021}. Thus fitting the CEF-phonon without the CEF-phonon coupling got close to the actual ground state, but to get the Hamiltonian quantitatively correct one must include the CEF-phonon coupling in the model. 

The fitted phonon energy is close to experiment, but the fitted CEF coupling is larger than DFT by an order of magnitude (mode 8 at 31~meV predicts $B_2^0=0.44$~meV, see Appendix \ref{app:details}), suggesting that the point-charge method underestimates the actual CEF-phonon coupling. 
It is not clear why, but 
the electronic coupling must be more dramatic than the nuclear displacements of the phonons suggest  \cite{Thalmeier_1984}. That said, the DFT+point-charge approximation still gave a meaningful estimate of the CEF coupling which was able to account for the anomalous third CEF mode.

\section{Conclusions}

We used density functional theory and point charge models to show a mechanism by which a low-energy phonon can couple to CEF modes at much higher energies, via a higher harmonic producing a splitting in the CEF mode energies. 
Although these extra modes could technically be considered vibronic bound states as they are produced by phonons, Figs. \ref{fig:Example} and \ref{fig:ToyModel} show that the split CEF modes have very little phonon character (judged by the phonon creation operators), and would instead appear as a CEF mode indistinguishable from any other CEF excited mode. 

The DFT+point-charge calculations dramatically underestimate the CEF-phonon coupling. But this has been noted in other contexts \cite{Thalmeier_1984} and is therefore not surprising, as it is an approximation to assume the electron orbitals interact precisely as the nuclear oscillation amplitudes. 
Nevertheless, they give a useful qualitative picture of the CEF-phonon coupling in KCeO$_2$, whereby a given phonon can be represented by its strongest Stevens operator term. 
With this  simplified model, we show how a CEF-phonon harmonic can explain extra CEF modes. 
Likely this effect is present in many other materials with other ions (e.g. NaYbSe$_2$ \cite{PhysRevB.103.035144}), but is most easily seen in Ce$^{3+}$ where a third mode glaringly contrasts with the simple single-ion model. 
(Whether the CEF-phonon coupling strength is ion-dependent is an interesting question, but outside the scope of this study.) 

We thus have solved the mystery of the extra mode observed in many Ce$^{3+}$ compounds. An extra CEF mode appears when an optical phonon creates a CEF-phonon harmonic which splits a high energy CEF eigenstate. 
This effect is generic, but is particularly pronounced in the $A$Ce$X_2$ compounds as the $B_2^1$ optical phonon strongly couples to the CEF Hamiltonian and matches the energy spacing of the CEF modes.

\acknowledgments

The work by AS was supported by the U.S. Department of Energy, Office of Basic Energy Sciences, Division of Materials Science and Engineering under project ``Quantum Fluctuations in Narrow-Band Systems.''
We acknowledge the support by the Institutional Computing Program at LANL, via the Center for Integrated Nanotechnologies, a DOE BES user facility, for computational resources.
SDW acknowledges support from the US Department of Energy (DOE), Office of Basic Energy Sciences, Division of Materials Sciences and Engineering under Grant No. DE-SC0017752. SJL gratefully acknowledges the support of the U.S. Department of Energy through the LANL/LDRD Program and the Center for Nonlinear Studies for this work.
We acknowledge helpful discussions with Filip Ronning, Priscila Rosa, and Mitchell Bordelon.

\appendix

\section{CEF-phonon calculations}\label{app:details}

\subsection{Calculated amplitude}

The density functional calculations report the phonon amplitudes in normalized units such that the sum of all atomic displacements squared per mode is 1 \cite{togo2023first}. That is, $\sum_j w_{j\nu}^2 = 1$ where $w_{j\nu}$ is the displacement of atom $j$ in phonon mode $\nu$. 
To compute this as a dimensionful displacement of a single phonon mode, we use the equation for the RMS amplitude $u$ of a simple harmonic oscillator with frequency $\omega_{\nu}$. 
\begin{equation}
	u_{j\nu} = \sqrt{\frac{\hbar}{2 \omega_{\nu} m_j}} (1+n)  w_{j\nu}
\end{equation}
where $m_j$ is the mass of atom $j$ and $n$ is the number of excited phonon modes ($n=1$ for calculating a single phonon mode displacement) \cite{togo2023first}. 

\subsection{Q-dependent calculated splitting}

\begin{figure*}
    \centering
    \includegraphics[width=0.95\textwidth]{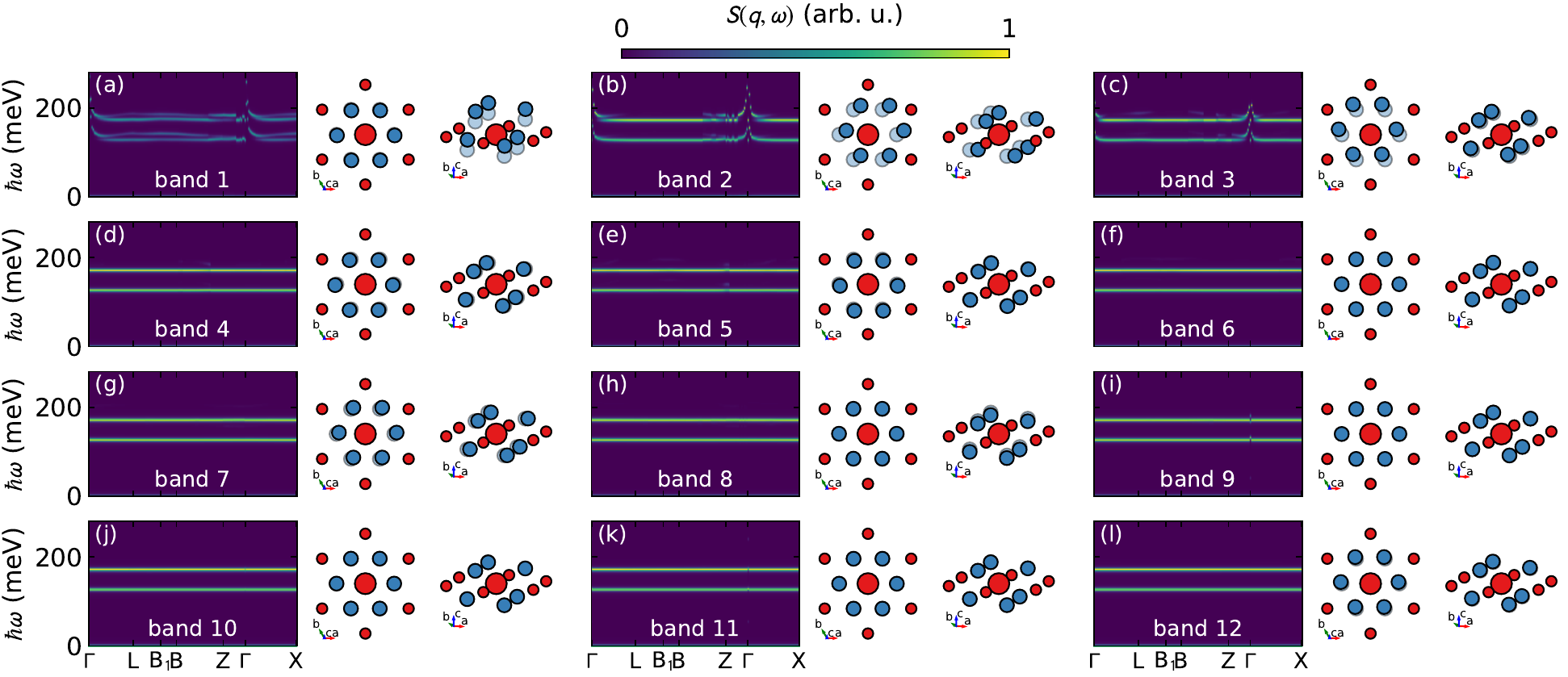}
    \caption{Calculated $\langle 0 | J_{\pm} | n \rangle$ of the KCeO$_2$ CEF-phonon eigenspectrum using the DFT calculated phonon distortions. The colormap shows the calculated intensity, and the schematic to the right of each panel illustrates the $Q=0$ phonon distortions. Blue circles indicate O atoms, red circles indicate Ce atoms (size indicates distance from the central ion; also note some distortions are barely visible, like bands 9-11). Note that the phonon modes are ordered in energy, which produces discontinuities in the eigenspectrum as a function of momentum when phonon modes cross in energy.}
    \label{fig:exhaustive_intensities}
\end{figure*}

\begin{figure*}
	\centering
	\includegraphics[width=0.95\textwidth]{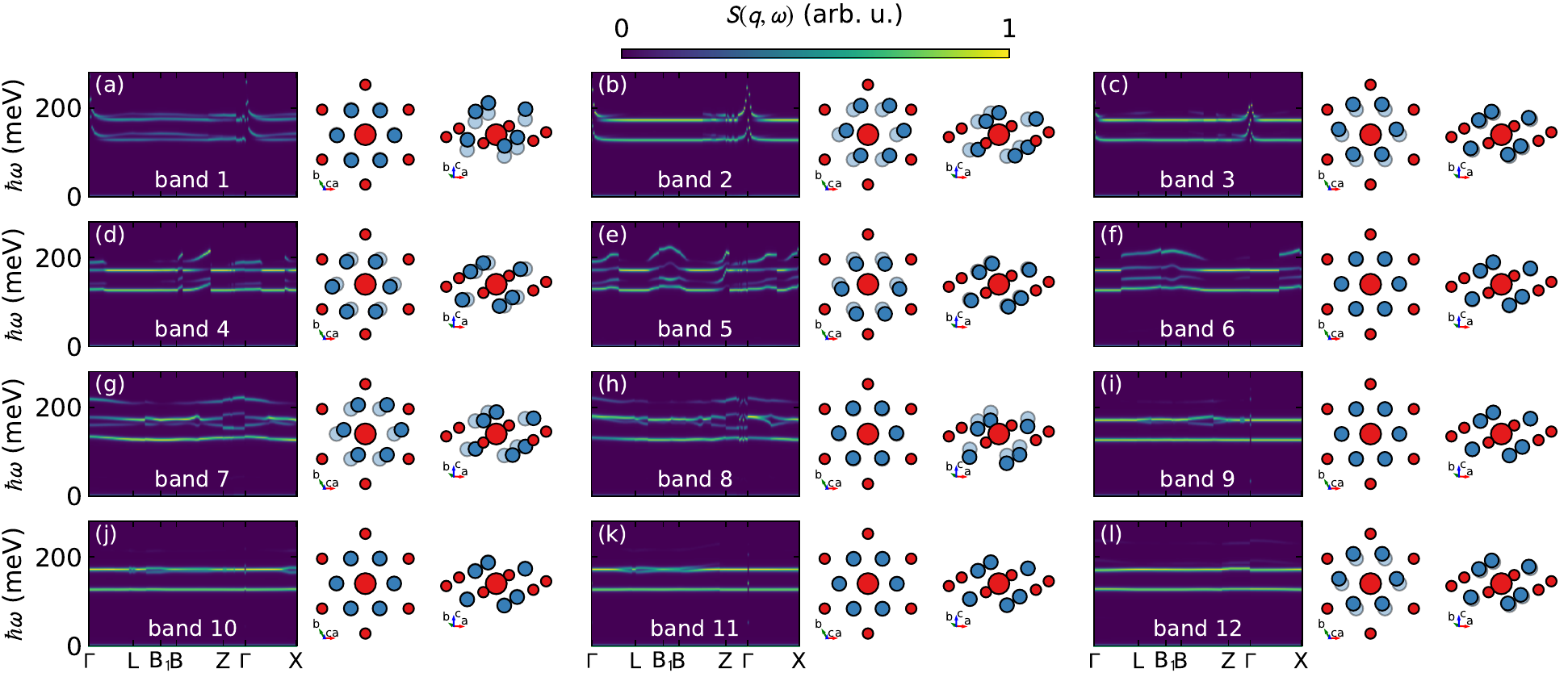}
	\caption{Calculated $\langle 0 | J_{\pm} | n \rangle$ of the KCeO$_2$ CEF-phonon eigenspectrum using the DFT calculated phonon distortions, but with the atomic displacements of modes 4-12 multiplied by 3. This modest increase in displacement produces noticeable results in the CEF spectrum.}
	\label{fig:exhaustive_intensities2}
\end{figure*}

Here we list the full results of the phonon-CEF diagonalization calculation. 
Figure \ref{fig:exhaustive_intensities} shows the momentum-dependent $\langle 0 | J_{\pm} | n \rangle$ showing the CEF intensities. Representative $\mathcal{O}_{\mu}$ for all phonons near $\Gamma$ are listed in Table \ref{tab:KCO_phononOperators}. 
For the most part, the phonon distortions in Fig. \ref{fig:exhaustive_intensities} have no effect on the CEF transitions. To amplify this effect, we multiply the phonon distortions of modes 4-12 by a factor of three and recalculated the CEF spectrum shown in Fig. \ref{fig:exhaustive_intensities2}. As shown by the new Stevens operator coefficients in Table \ref{tab:KCO_phononOperators}, this increases the coupling by an order of magnitude. 
Although the reason for the point-charge model's underestimate is unknown, when the coupling is increased by this amount it begins to reproduce the multiple CEF peaks observed in KCeO$_2$. 

\begin{table*}
	\caption{CEF operators for ${\bf Q} = (0.02,0,0)$ phonons in KCeO$_2$ calculated from DFT dispacelements via the point charge model. Modes 7-8 produce a large splitting of the CEF eigenstates with extra modes appearing at high energy, which are dominated by a $B_2^1$ distortion.}
	\begin{ruledtabular}
		\begin{tabular}{c|c|cccccccc}
Band  &  $\hbar \omega_{phonon}$ (meV)  &  $B_2^0$ & $B_2^1$ & $B_2^2$ & $B_4^0$ & $B_4^1$ & $B_4^2$ & $B_4^3$ & $B_4^4$ \\
\hline
1  &  0.62  &  -0.86356 & -0.655678 & -0.485606 & 0.116055 & -0.019869 & -0.025409 & -2.181088 & -0.025966 \\
2  &  0.95  &  -0.765146 & -3.879517 & -0.132704 & 0.008894 & 0.071082 & -0.041184 & -1.576603 & 0.090931 \\
3  &  1.66  &  -0.435758 & 2.497297 & 0.328281 & 0.004841 & -0.029888 & 0.023324 & -0.895509 & -0.047835 \\
4  &  12.5  &  -0.030139 & -0.156557 & -0.006892 & 0.000289 & 0.002671 & -0.001172 & -0.06265 & 0.004565 \\
5  &  12.6  &  -0.036535 & 0.190866 & 0.009624 & 0.000351 & -0.003194 & 0.001425 & -0.07593 & -0.005476 \\
6  &  23.97  &  -0.000437 & 0.00467 & 0.00197 & 5.1$\times 10^{-5}$ & 6.7$\times 10^{-5}$ & 0.000106 & -0.001437 & 4.1$\times 10^{-5}$ \\
7  &  30.26  &  -0.088487 & -0.458557 & -0.019746 & 0.000863 & 0.007868 & -0.003549 & -0.183693 & 0.013155 \\
8  &  38.51  &  -0.021847 & -0.065706 & -0.0472 & 0.009678 & -0.002506 & -0.002304 & -0.148166 & -0.002539 \\
9  &  43.72  &  -3$\times 10^{-6}$ & -1.4$\times 10^{-5}$ & -1$\times 10^{-6}$ & 0.0 & 0.0 & -0.0 & -6$\times 10^{-6}$ & 0.0 \\
10  &  43.74  &  -0.000113 & -0.000673 & -0.000577 & 4$\times 10^{-5}$ & -3.4$\times 10^{-5}$ & -2.7$\times 10^{-5}$ & -0.000672 & -3.6$\times 10^{-5}$ \\
11  &  54.77  &  -4.5$\times 10^{-5}$ & 0.000335 & 8$\times 10^{-5}$ & 1$\times 10^{-6}$ & -0.0 & 5$\times 10^{-6}$ & -0.0001 & -3$\times 10^{-6}$ \\
12  &  62.41  &  -0.040934 & 0.25504 & 0.038476 & 0.000484 & -0.002136 & 0.002937 & -0.086085 & -0.00472 \\
	\end{tabular}\end{ruledtabular}
	\label{tab:KCO_phononOperators}
\end{table*}

\begin{table*}
	\caption{CEF operators for ${\bf Q} = (0.02,0,0)$ phonons in KCeO$_2$ calculated from DFT displacements via the point charge model (same as Table \ref{tab:KCO_phononOperators}) but with modes 4-12 displacements multiplied by 3. These values are much closer to the fitted CEF-phonon operators.}
	\begin{ruledtabular}
		\begin{tabular}{c|c|cccccccc}
Band  &  $\hbar \omega_{phonon}$ (meV)  &  $B_2^0$ & $B_2^1$ & $B_2^2$ & $B_4^0$ & $B_4^1$ & $B_4^2$ & $B_4^3$ & $B_4^4$ \\
\hline
1  &  0.62  &  -22.120211 & -3.995187 & -1.300151 & 0.104792 & -0.041796 & 0.017506 & -9.868925 & -0.095494 \\
2  &  0.95  &  -2.751859 & -20.364184 & -3.326117 & 0.098785 & 0.442707 & -0.442928 & -9.086257 & 0.065405 \\
3  &  1.66  &  -3.954439 & 15.364462 & 3.584138 & 0.076431 & -0.304765 & 0.094928 & -6.226998 & -0.057675 \\
4  &  12.5  &  -0.269984 & -1.389464 & -0.055962 & 0.002766 & 0.024275 & -0.011846 & -0.558529 & 0.037847 \\
5  &  12.6  &  -0.327578 & 1.669838 & 0.102463 & 0.003409 & -0.02933 & 0.011587 & -0.674676 & -0.04449 \\
6  &  23.97  &  -0.003933 & 0.042001 & 0.017724 & 0.00046 & 0.0006 & 0.000952 & -0.012927 & 0.000368 \\
7  &  30.26  &  -0.780466 & -3.955734 & -0.134713 & 0.009107 & 0.072583 & -0.042245 & -1.608173 & 0.092264 \\
8  &  38.51  &  -0.380171 & -0.540642 & -0.400416 & 0.076294 & -0.018511 & -0.020589 & -1.313854 & -0.021753 \\
9  &  43.72  &  -2.5$\times 10^{-5}$ & -0.000129 & -6$\times 10^{-6}$ & 0.0 & 2$\times 10^{-6}$ & -1$\times 10^{-6}$ & -5.2$\times 10^{-5}$ & 4$\times 10^{-6}$ \\
10  &  43.74  &  -0.001022 & -0.006058 & -0.005188 & 0.000359 & -0.000305 & -0.000244 & -0.006049 & -0.00032 \\
11  &  54.77  &  -0.000407 & 0.003017 & 0.000716 & 8$\times 10^{-6}$ & -3$\times 10^{-6}$ & 4.6$\times 10^{-5}$ & -0.000898 & -3$\times 10^{-5}$ \\
12  &  62.41  &  -0.36518 & 2.236982 & 0.356552 & 0.004399 & -0.02043 & 0.023884 & -0.757653 & -0.03707 \\
	\end{tabular}\end{ruledtabular}
	\label{tab:KCO_phononOperators2}
\end{table*}

\subsection{Fitted CEF eigenspectrum}

The fitted CEF eigenspectrum of KCeO$_2$ of the model shown in Fig. \ref{fig:refit} is shown in Table \ref{tab:NewFitEigenvectors}. Note that these CEF eigenstates are what they would be in the absence of phonon coupling (``raw CEF model'' in Fig. \ref{fig:refit}).  

\begin{table}
	\caption{Eigenvalues (in meV) and eigenvectors with uncertainty of the best fit KCeO$_2$ Hamiltonian including the phonon coupling.}
	\begin{ruledtabular}
		\begin{tabular}{c|cccccc}
			$\hbar \omega$ &$| -\frac{5}{2}\rangle$ & $| -\frac{3}{2}\rangle$ & $| -\frac{1}{2}\rangle$ & $| \frac{1}{2}\rangle$ & $| \frac{3}{2}\rangle$ & $| \frac{5}{2}\rangle$ \tabularnewline
			\hline 
			0.0 & 0.0 & 0.0 & -0.919(12) & 0.0 & 0.0 & 0.39(3) \tabularnewline
			0.0 & -0.39(3) & 0.0 & 0.0 & -0.919(12) & 0.0 & 0.0 \tabularnewline
			118 & 0.0 & -1.0 & 0.0 & 0.0 & 0.0 & 0.0 \tabularnewline
			118 & 0.0 & 0.0 & 0.0 & 0.0 & -1.0 & 0.0 \tabularnewline
			162 & -0.919(12) & 0.0 & 0.0 & 0.39(3) & 0.0 & 0.0 \tabularnewline
			162 & 0.0 & 0.0 & -0.39(3) & 0.0 & 0.0 & -0.919(12) \tabularnewline
	\end{tabular}\end{ruledtabular}
	\label{tab:NewFitEigenvectors}
\end{table}

%


\end{document}